\documentclass{article}
\usepackage{srcltx}
\usepackage[T2A]{fontenc}
\usepackage[utf8]{inputenc}
\usepackage[english]{babel}

\usepackage[a4paper,top=4cm,bottom=3.5cm,left=2.5cm,right=2.5cm,marginparwidth=1.75cm]{geometry}
\usepackage{graphicx}

\begin{document}

\title{WHITE DWARFS IN A UNIFORM SPHERE APPROXIMATION, WITH GENERAL RELATIVITY EFFECTS TAKEN INTO ACCOUNT}

\author{G. S. Bisnovatyi-Kogan\footnote{{Institute of Space Research, Russian Academy of Sciences, Moscow.} {Moscow Physics-Technical Institute MFTI, Dolgoprudnyi.}},
Е.А. Патраман \footnote{
{Moscow Physics-Technical Institute MFTI, Dolgoprudnyi.}{Институт космических исследований РАН, Москва,}}}

\maketitle

\begin{abstract}

\bigskip

 The limiting mass of cold white dwarfs was first calculated by E. Stoner in an approximate model of a
uniform star and was soon reduced by ~20
an exact solution of the equations for the stellar equilibrium. Here we examine uniform models of white
dwarfs taking general relativity effects and the influence of finite temperature into account. Solutions
are obtained in the form of finite analytic formulas and, for masses differing by no more than ~20
the exact solutions, found by numerical integration of the differential equations for the stellar equilibrium.
\end{abstract}


 \medskip

\section{Introduction}
In studies of the structure of white dwarfs it is usually observed that they can be in equilibrium only for masses
not exceeding some limit known as the Chandrasekhar limit. For a carbon-oxygen chemical composition, with two
nucleons per electron, i.e., $\mu_{e} = 2$, this limit equals 
$\approx 1.46M_{\odot}$. Stoner \cite{stoner30}, who examined a model of a white dwarf
with a uniform density (see \cite{thomas}, as well), concluded for the first time that there is an upper bound on the mass
for cold stars whose equilibrium is maintained by the pressure of degenerate electrons. He generalized the earlier discussion on degenerate electrons from \cite{fowler26,frenkel28} to the case of high density under conditions of ultrarelativistic
degeneracy in which the equation of state for cold matter takes the form \cite{bk89}

\begin{equation}\label{eq1}
    P(\rho) = \frac{(3\pi^2)^{1/3}}{4}\frac{\hbar c}{(\mu_e m_u)^{4/3}}\rho^{4/3} = K\rho^{4/3}
\end{equation}
Here $\mathrm{\mu_{e}}$ is the number of nucleons per electron,
  $\mathrm{m_{u}}$  is the atomic mass unit, equal to 1/12 of
   the mass of the isotope $\mathrm{^{12}C}$.

The mass of a polytropic star, corresponding to $\mathrm{\gamma} = 4/3$, n = 3,  according to the Emden theory
 is independent of the density and is uniquely determined by
  the parameter K in the form \cite{bk89}

\begin{equation}\label{eq2}
    M_p = 4\pi \Big(\frac{K}{\pi G}\Big)^{3/2} \cdot M_3,\quad M_3 = 2.018
\end{equation}
Using (\ref{eq2}), Chandrasekhar \cite{chandra31} and Landau \cite{landau32},independently and almost simultaneously, obtained a limiting mass
for a white dwarf with the equation of state(\ref{eq2}) in the form 

\begin{equation}\label{eq3}
    M_{wd} = \frac{\sqrt{3\pi}}{2}\Big(\frac{\hbar c}{G}\Big)^{3/2} \frac{M_3}{(\mu_e m_u)^2} = \frac{5.83}{\mu_e^2} M_{\odot}
\end{equation}
To determine the limiting mass of observed white dwarfs, Chandrasekhar \cite{chandra31}, following Stoner \cite{stoner30}, obtained a value $\mathrm{M_{wd}} = 0.933 \mathrm{M_{\odot}}$.
 This refined Stoner’s value $\mathrm{M_{wd}} = 1.1
  \mathrm{M_{\odot}}$ for the same value of $\mathrm{\mu_{e}}
   = 2.5$ in
the uniform density model. It follows from the theory of stellar evolution, as well as from observations, that almost
all white dwarfs consist of a mixture of carbon $\mathrm{^{12}C}$ and oxygen  $\mathrm{^{16}O}$ for which $\mathrm{\mu_{e}} = 2$ \cite{shatzman}, while $\mathrm{M_{wd}} = 1.46 \mathrm{M_{\odot}}$. For
the first time a realistic value of the mass limit for white dwarfs was obtained in \cite{landau32}, which merits a more correct designation as the Stoner-
Chandrasekhar-Landau limit.

We note that the values of the limiting masses of white dwarfs are given here based on refined modern values
for all the constants, which has led to a difference of a few percent from the values given in the original papers.
In this paper a uniform model is used to construct approximate models of white dwarfs with arbitrary masses
at a finite temperature taking into account the post-Newtonian corrections to Newtonian gravitation owing to the
effects of the general theory of relativity (GR). A comparison of the results for the limiting masses of white dwarfs
in the exact and uniform models shows that the errors in determining all the values in the uniform model are around
20\%.

\section{White dwarfs in the approximation of a sphere of constant density}

The energy method used by Stoner \cite{stoner30} for white dwarfs and later for the general case in \cite{zn65, bk66}. The density distribution is taken to be specified, with a single
parameter in the form of the central density, with variations in which the star changes homologically.
For uniform stars this kind of parameter is the constant density over the star. We write the total energy 
$\mathrm{\varepsilon}$ of
a uniform star in the form

\begin{equation}\label{eq4}
    \varepsilon=E_T M + \varepsilon_G = E_TM - \frac{3}{5} G\frac{M^2}{R} = E_TM - \frac{3}{5}G\Big(\frac{4\pi M^5}{3}\Big)^{1/3}\rho^{1/3}
\end{equation}
Here $\mathrm{E_{T}}$ is the internal energy per unit mass and the gravitational energy of the uniform sphere  $\mathrm{\varepsilon_{G}}$ is defined
in \cite{lltp}. The equilibrium state and the stability condition for a star with mass $M$ are thus determined, respectively, by the equations

\begin{equation}\label{eq5}
    0 = \frac{d\varepsilon}{d\rho^{1/3}} = 3MP\rho^{-4/3} - \frac{3}{5}G\Big(\frac{4\pi M^5}{3}\Big)^{1/3}; \quad \frac{d^2\varepsilon}{d(\rho^{1/3})^2} = 9MP\rho^{-5/3}\Big(\gamma-\frac{4}{3}\Big) > 0
\end{equation}
Here we have used the thermodynamic relations \cite{llsp} 

\begin{equation}\label{eq6}
    P = \rho^2\Big(\frac{\partial E_T}{\partial \rho}\Big)_{|S} ;\quad \gamma=\Big(\frac{\partial \ln P}{\partial \ln\rho}\Big)_{|S}
\end{equation}
The first equilibrium relation in (\ref{eq5}) yields the only solution for the mass of a uniform star $\mathrm{M_{u}}$, in the case
of the equation of state $P = K \rho^{4/3}$, in the form

\begin{equation}\label{eq7}
    M_u =5.463 \Big(\frac{K}{G}\Big)^{3/2}
\end{equation}
On comparing the masses $\mathrm{M_{u}}$ from (\ref{eq7})) and $\mathrm{M_{p}}$ from (\ref{eq3}), we obtain for the limiting mass

\begin{equation}\label{eq8}
    M_u = \Big(\frac{5}{4}\Big)^{3/2}\frac{\sqrt{3}}{M_3}M_{wd} = 1.199 M_{wd}
\end{equation}
 Thus, the approximate value for the limiting mass of a uniform star according to Stoner is roughly 20\% greater than
the exact value. For cold white dwarfs we have the following equation of state \cite{bk89}
 
\begin{equation}\label{eq10}
    P_{ec} = \frac{m_e^4 c^5}{24 \pi^2 \hbar^3}[y(2y^2-3)\sqrt{y^2+1}+3\sinh^{-1}y], \quad y = \Big(\frac{3\pi^2\rho}{\mu_em_u}\Big)^{1/3}\frac{\hbar}{m_e c} = \Big(\frac{1.027\rho}{10^9\mu_e}\Big)^{1/3}.
\end{equation}
For cold white dwarfs at arbitrary density we obtain, with account of (\ref{eq10}), the following expression for the dependence of the stellar mass on its density  $M_{uc}(\rho)$ in the form

\begin{equation}\label{eq9}
    M_{uc} = 5.463\Big(\frac{P_{ec}\rho^{-4/3}}{G}\Big)^{3/2}.
\end{equation}
When the small corrections owing to deviations from an ultrarelativistic gas for $y \gg 1$,
 and  small temperature corrections  $\alpha = \frac{m_e c^2}{kT} \gg 1$,
are taken into account, the equation of state takes
the form \cite{bk89}

\begin{equation}\label{eq11}
    P = \frac{m_e^4 c^5}{12 \pi^2 \hbar^3} y^4 \Big(1 - \frac{1}{y^2} + \frac{2\pi^2}{3\alpha^2y^2}\Big) = P(\rho) = \frac{(3\pi^2)^{1/3}}{4}\frac{\hbar c}{(\mu_e m_u)^{4/3}}\rho^{4/3} \Big(1 - \frac{1}{y^2}+\frac{2\pi^2}{3\alpha^2y^2}\Big)
\end{equation}
Then in the uniform model the dependence of the mass on the density for high densities, taking (\ref{eq9}) into account,
is written in the form

\begin{equation}\label{eq12}
    M_u = 5^{3/2}\frac{3\sqrt{\pi}}{16}(\frac{\hbar c}{G})^{3/2}\frac{1}{(\mu_e m_u)^2}\Big(1 - \frac{1}{y^2}+\frac{2\pi^2}{3\alpha^2y^2}\Big)^\frac{3}{2}
\end{equation}
For white dwarfs at arbitrary density, with small temperature corrections, the equation of state takes
the form \cite{bk89}

\begin{equation}\label{eq11a}
P(\rho) = \frac{m_e^4 c^5}{24 \pi^2 \hbar^3} 
\Big[y(2y^2-3)\sqrt{y^2+1}+3\sinh^{-1}y
+ \frac{4\pi^2}{3\alpha^2}y\frac{y^2+2}{\sqrt{y^2+1}}\Big].
\end{equation}
The white dwarf mass $M_{ut}$ in the uniform model, at arbitrary density, with small temperature corrections, is defined as

\begin{equation}\label{eq12а}
M_{ut}= 5^{3/2}\frac{3\sqrt{\pi}}{16}(\frac{\hbar c}{G})^{3/2}\frac{1}{(\mu_e m_u)^2}\Big\{\frac{1}{2y^4}
\Big[y(2y^2-3)\sqrt{y^2+1}+3\sinh^{-1}y
 + \frac{4\pi^2}{3\alpha^2}y\frac{y^2+2}{\sqrt{y^2+1}}
 \Big]\Big\}^{3/2}.
\end{equation}

\section{Models of uniform white dwarfs with account of small corrections for GR}

\begin{figure}[b]
\begin{center}
\includegraphics[width=0.8\linewidth]{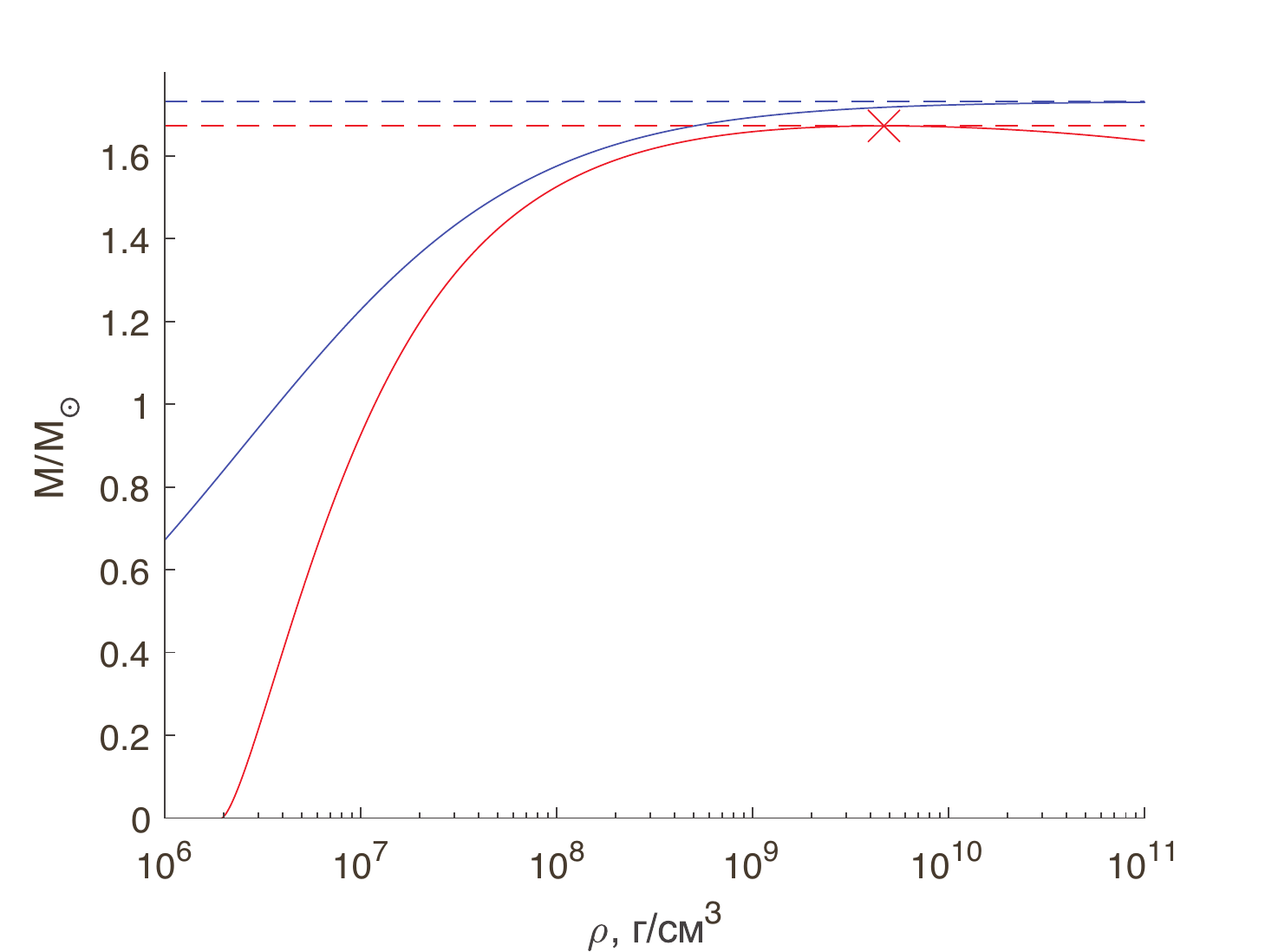}
\caption{The dependence of the mass on the density around maximal mass, for cold WD, with GR
effects neglected and included. The blue curve
corresponds to a uniform model with GR effects
neglected, with equation of state of (\ref{eq11}). The red curve shows a uniform model,
using (\ref{eq11})), with small corrections for GR included.
The maximum is reached at the point
$\rho_m = 4.689 \cdot10^9$ г/см$^3$, $M_m = 1.672 M_\odot$.} 
\label{fig1}
\end{center}
\end{figure}

\begin{figure}[b]
\begin{center}
\includegraphics[width=0.8\linewidth]{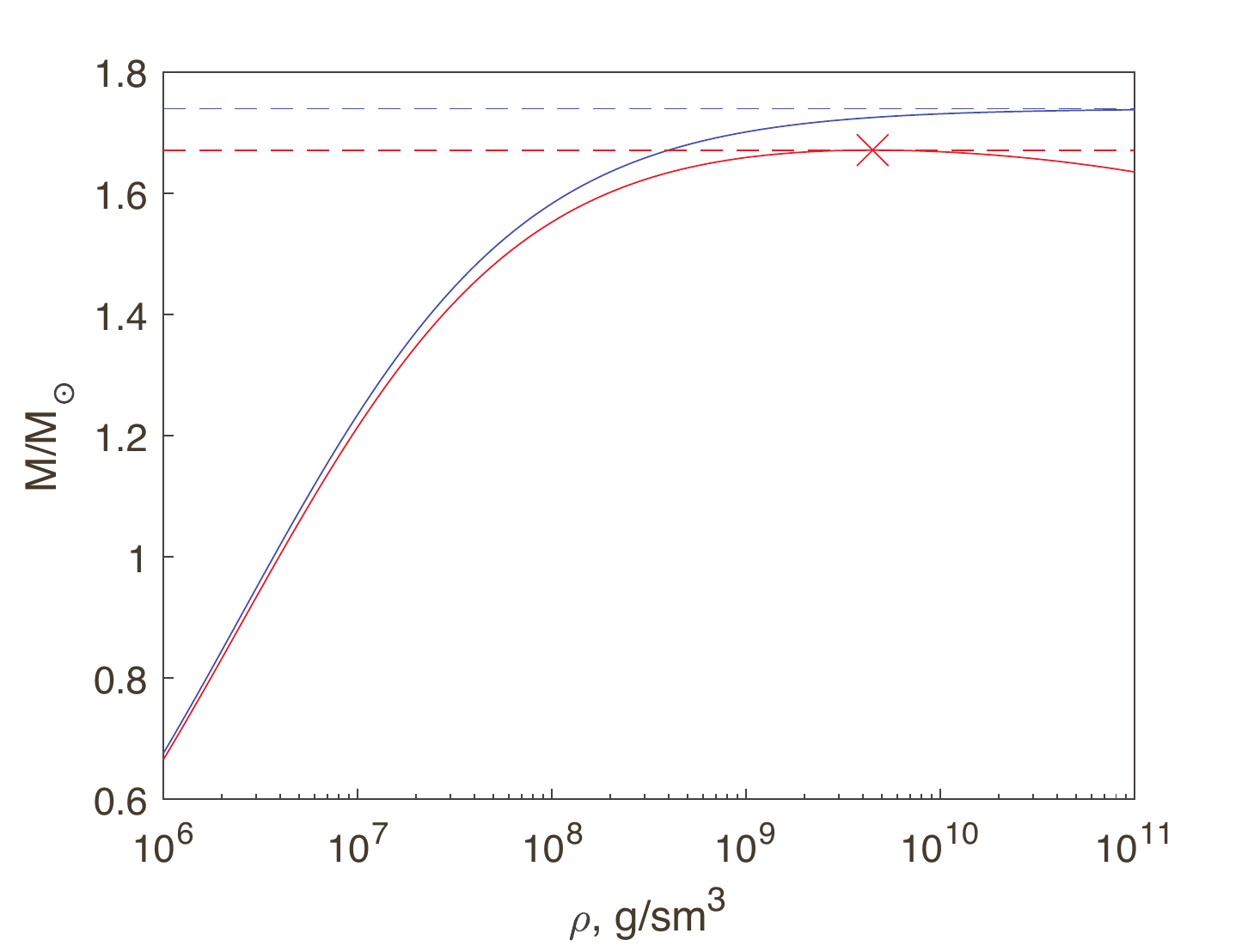}
\caption{The dependence of the mass at arbitrary density, for cold WD with GR
effects neglected and included. The blue curve
corresponds to a uniform model with GR effects
neglected, with equation of state of (\ref{eq11a}). The red curve shows a uniform model,
using (\ref{eq11a})), with small corrections for GR included.
The maximum is reached at the point
$\rho_m = 4.689 \cdot10^9$ г/см$^3$, $M_m = 1.672 M_\odot$.}
\label{fig1a}
\end{center}
\end{figure}

We now examine models of uniform white dwarfs taking into
account the post-Newtonian corrections for the GR.
For equilibrium stars the post-Newtonian corrections to the energy are defined in \cite{zn65}, see also \cite{zn71}, in
the integral form, which are calculated analytically for a uniform sphere as

\begin{eqnarray}\label{eq13}
\Delta E = I_1 + I_2 + I_3 + I_4 + I_5,\qquad\qquad\qquad\qquad\\
    I_1 = -\frac{G}{c^2}E_{T}\int m\frac{dm}{r} = 
-\frac{3}{5}\Big(\frac{4\pi}{3}\Big)^{1/3}\frac{G}{c^2}M^{5/3}
E_T\rho^{1/3}=    
    - \frac{9}{25} \frac{G^2}{c^2}\Big(\frac{4 \pi \rho}{3}\Big)^\frac{2}{3} M^\frac{7}{3},\nonumber\\
    I_2 = -\frac{G^2}{2c^2}\int m^2\frac{dm}{r^2} = -(3)^\frac{1}{3} \frac{G^2}{14 c^2}(4 \pi \rho)^\frac{2}{3} M^\frac{7}{3},\qquad\qquad \nonumber\\
    I_3 = -\frac{G}{c^2}\int\Big(\int E_{T} dm\Big)\frac{dm}{r} =-\frac{3}{5}\Big(\frac{4\pi}{3}\Big)^{1/3}\frac{G}{c^2}M^{5/3}
E_T\rho^{1/3}=  
        - \frac{9}{25} \frac{G^2}{c^2}\Big(\frac{4 \pi \rho}{3}\Big)^\frac{2}{3} M^\frac{7}{3},\nonumber\\
    I_4 = \frac{G^2}{c^2}\int\Big(\int \frac{m dm}{r}\Big)\frac{dm}{r} = 3^\frac{4}{3} \frac{G^2}{35 c^2}(4 \pi \rho)^\frac{2}{3} M^\frac{7}{3},\qquad\qquad \nonumber\\
    I_5 = -\frac{G^2}{c^2}\int\Big(\int m r dr\Big)\frac{m dm}{r^4} = -3^\frac{1}{3} \frac{G^2}{35 c^2}(4 \pi \rho)^\frac{2}{3} M^\frac{7}{3}. \qquad\qquad \nonumber
\end{eqnarray}
Ultimately we obtain the correction for the GR for a uniform sphere in the form

\begin{equation}\label{eq14}
    \Delta E = -\frac{6}{5}\Big(\frac{4\pi}{3}\Big)^{1/3}\frac{G}{c^2}M^{5/3}E_T\rho^{1/3} - \frac{3}{70}\Big(\frac{4\pi}{3}\Big)^{2/3} M^{7/3}\rho^{2/3}
\end{equation}
For a polytropic sphere the following relation holds, derived from the virial theorem \cite{bk89}

\begin{equation}\label{eq15}
    \varepsilon_T=E_TM=-\frac{n}{3}\varepsilon_G=\frac{n}{5}G\Big(\frac{4\pi}{3}\Big)^{1/3}M^{5/3}\rho^{1/3}
\end{equation}
The relation (\ref{eq15}) is used in (\ref{eq13}) for $n=3$, and in the first term of (\ref{eq15}), leading to the expression for the GR correction for a polytropic star with
uniform density and $n=3$ in the form

\begin{equation}\label{eq16}
    \Delta E= -1.982\frac{G^2}{c^2}M^{7/3}\rho^{2/3}
\end{equation}
For the Emden polytropic model with n = 3 the GR corrections are written as \cite{zn65}

\begin{equation}\label{eq17}
    \Delta E = -0.93\frac{G^2}{c^2}M^{7/3}\rho^{2/3}
\end{equation}
When the corrections for the GR are taken into account, the approximate equation for the equilibrium in a uniform
model including (\ref{eq5}) и (\ref{eq16}) is written in the form

\begin{eqnarray}\label{eq18}
    0 = \frac{d\varepsilon}{d\rho^{1/3}} = 3MP\rho^{-4/3} - \frac{3}{5}G\Big(\frac{4\pi M^5}{3}\Big)^{1/3} + \frac{d\Delta E}{d\rho^{1/3}} =\nonumber\\
    =3MP\rho^{-4/3} - \frac{3}{5}G\Big(\frac{4\pi M^5}{3}\Big)^{1/3} - 3.964\frac{G^2}{c^2}M^{7/3}\rho^{1/3}
\end{eqnarray}
The GR correction is calculated here using the virial theorem for the uniform sphere at  $n=3$, corresponding to the white dwarf which mass close to the maximal one. For the low mass white dwarfs GR correction is negligibly small, so we may use it in the form (\ref{eq16}) for any lower mass as well.
The dependence  $M(\rho)$ for a uniform white dwarf model including GR effects following from (\ref{eq18}), with the
equation of state (\ref{eq11}) is shown in Fig. \ref{fig1}. The maximum mass is reached at $\rho_m = 4.689 \cdot10^9$ g/sm$^3$, $M_m = 1.672 M_\odot$.
Analogous curves for white dwarfs in exact polytropic models are constructed in \cite{bk66}. 
The dependence  $M(\rho)$ for a uniform white dwarf model including GR effects following from (\ref{eq18}), with the
equation of state (\ref{eq11a}), is shown in Fig. \ref{fig1a}.
  
\begin{figure}[h]
\center{\includegraphics[width=0.8\linewidth]{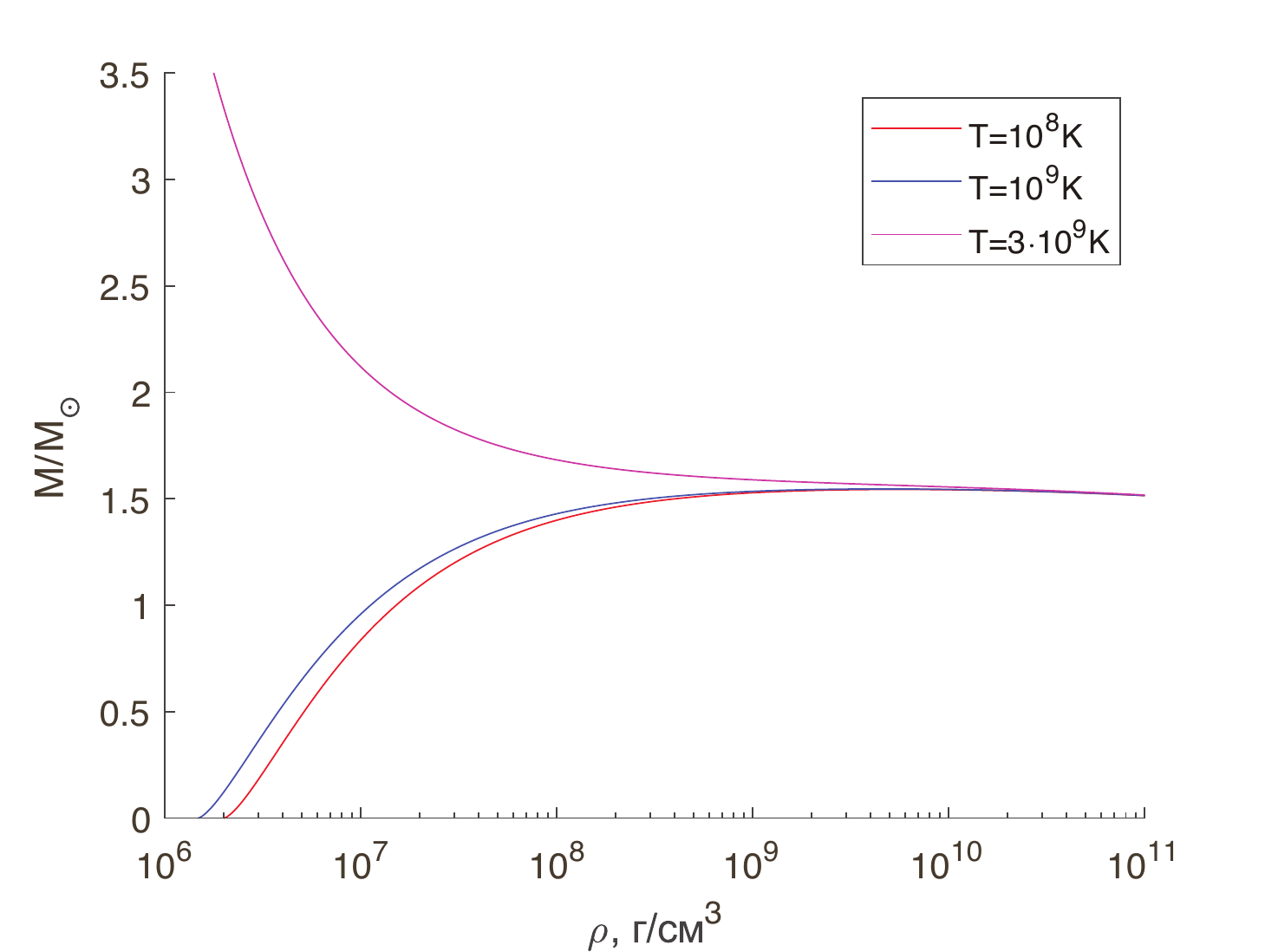}}
\caption{Plots for isothermal uniform white dwarfs at high
densities around the mass maximum, with GR effects taken into account, according to (\ref{eq11}),(\ref{eq19}).}
\label{fig2}
\end{figure}

\begin{figure}[h]
\center{
\includegraphics[width=0.8\linewidth]{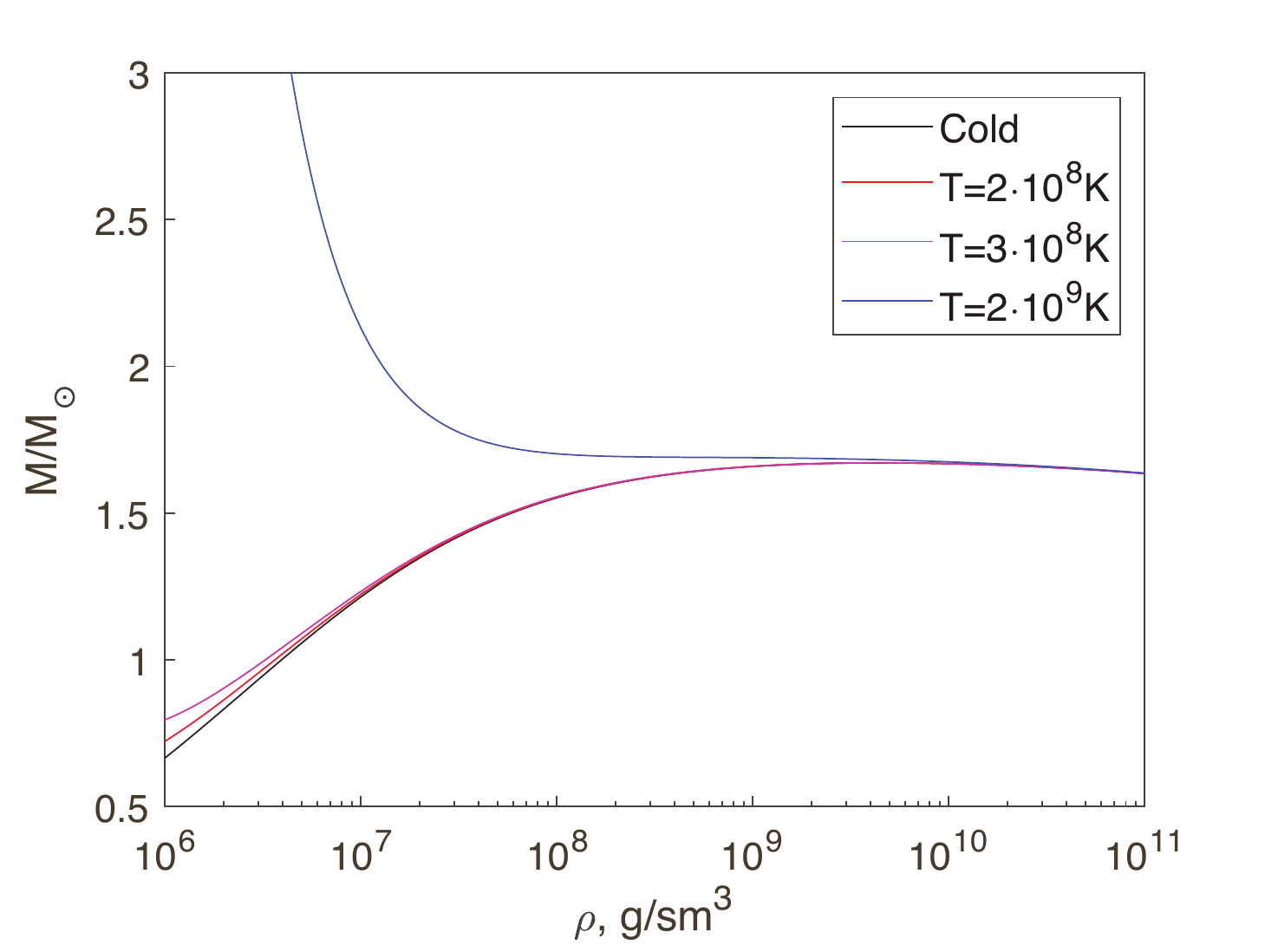}
\caption{Plots for isothermal uniform white dwarfs at arbitrary
densities, with GR effects taken into account, according to (\ref{eq11a}),(\ref{eq19}).}}
\label{fig2a}
\end{figure}

A comparison shows that the curves for the uniform models around mass maximum are roughly 20
than for the corresponding curves from \cite{bk66}, and lose stability owing to GR effects at a density of roughly a factor
of 5 lower because of the greater influence of these effects in the uniform model compared to the exact polytropic
model with n = 3. Using Eq. (\ref{eq18}) we obtain an equation for the mass of a uniform white dwarf with GR effects taken into account at a finite temperature:

\begin{equation}\label{eq19}
    M = \left( \frac{-\frac{3}{5}G\left(\frac{4\pi}{3}\right)^\frac{1}{3} + \sqrt{ (\frac{3}{5}G\left(\frac{4\pi}{3}\right)^\frac{1}{3})^2 + 47.568\frac{G^2}{c^2}\rho^{-1}P}}{7.928\frac{G^2}{c^2}\rho^\frac{1}{3}} \right)^\frac{3}{2}
\end{equation}
The dependence of the mass on the density of uniform isothermal white dwarfs for different temperatures with GR effects
taken into account, from (\ref{eq19}),  is shown in Fig. \ref{fig2} for the  equation of state (\ref{eq11}). Similar curves for isothermal white dwarfs in the exact polytropic model have been obtained in \cite{bk66}. 
The dependence of the mass in the same model at arbitrary density, for the  equation of state (\ref{eq11a}),  is shown in Fig.~ \ref{fig2a}.

\noindent 
Note, that the influence of GR effects on the stability of white dwarfs was first studied by Kaplan \cite{kap49}.

\section*{Acknowledgements}

This work was partially supported by RFFI grant 20-02-00455.

\end{document}